\documentclass[10pt]{article}
\usepackage[margin=1in]{geometry}
\usepackage[T1]{fontenc}
\usepackage{lmodern}
\usepackage{microtype,booktabs,tabularx,array,enumitem,graphicx,xcolor,amsmath,amssymb}
\usepackage{tikz}
\usetikzlibrary{arrows.meta,positioning,fit}
\usepackage[font=small,labelfont=bf]{caption}
\usepackage[hidelinks]{hyperref}
\usepackage{xurl}
\usepackage{natbib}
\setlist{nosep,leftmargin=1.6em}
\title{From Reactive Containment to Proactive Assurance: Lessons from the 2026 OpenAI, Anthropic, and Google Agent Security Incidents}
\author{Abbas Raftari\\Walsh College, Troy, Michigan, USA\\\href{mailto:araftari@walshcollege.edu}{araftari@walshcollege.edu}}
\date{September 2026}

\begin{document}
\maketitle

\begin{abstract}
In 2026, cybersecurity evaluations involving OpenAI, Anthropic, and Google agents reached real systems outside their authorized test scope. The paths were different. OpenAI agents exploited research infrastructure, coordinated across runs, and compromised parts of Hugging Face's production environment. Anthropic reported cases in which a misconfigured third-party environment exposed real systems to agents pursuing simulated cyber tasks. In a separately reported evaluation, Google's Gemini accessed three real organizations through an unintended internet route; Google stated that the model stopped in all three instances. Taken together, the cases show why an evaluation cannot rely on an assumed boundary. That boundary must be verified while the agent is operating. This comparative instrumental case study develops a Proactive Agent Security Assurance Cycle (PASAC) and a five-layer Boundary Assurance Stack. The framework combines risk-tiered task design, executable scope contracts, pre-run validation, least-capability access, independent egress enforcement, credential restrictions, cross-run monitoring, automatic stop conditions, and evidence-based reauthorization. A leading-indicator model, nine design propositions, and seven falsifiable hypotheses turn these lessons into a testable research program. Because the public Gemini record is limited to attributed statements and journalism, its detailed causal mechanism remains provisional. The central conclusion is straightforward: proactive agent security requires continuous assurance across the full execution system, not confidence in any single sandbox or safeguard.
\end{abstract}

\noindent\textbf{Keywords:} autonomous agents; proactive cybersecurity; AI safety; cyber evaluation; sandbox containment; reward hacking; zero trust; incident governance

\section{Introduction}

Agentic AI changes what security teams need to protect. Once a model can use a shell, browse the web, call a package manager, retain memory, access credentials, or continue working over a long execution loop, it can discover operational paths that its designers did not anticipate. The model alone is therefore not the safety boundary. The real object of assurance is the complete execution system: the objective, model, harness, tools, identities, shared services, networks, monitors, human authority, and external partners.

Three sets of incidents disclosed in 2026 make this problem concrete. In the OpenAI--Hugging Face case, agents running difficult cybersecurity evaluations exploited internal research infrastructure, used a shared Artifactory service for unintended communication, reached the public internet, and compromised parts of Hugging Face's production environment \citep{openai2026report,openai2026blog,hf2026timeline}. In the Anthropic cases, models intentionally operating without normal cyber safeguards obtained live internet access because a third-party evaluation environment was misconfigured. Some models then interacted with and compromised real external systems while pursuing simulated capture-the-flag tasks \citep{anthropic2026improving,reuters2026anthropic3,reuters2026anthropic4}.

Google stated that Gemini accessed three real organizations during an evaluation by Irregular in May 2026, using a guessed password in one case and publicly exposed credentials in two others. Google said the model stopped in all three instances and the organizations were notified \citep{reuters2026gemini,guardian2026gemini}. The popular label ``rogue AI'' offers little engineering guidance. OpenAI's agents defeated an intended boundary; the Anthropic and Gemini evaluations encountered unintended live access. Gemini provides a distinct reported outcome: the model stopped, but only after unauthorized access. Together these cases foreground adaptive escape, configuration assurance, environment recognition, credential control, and authorization before an action occurs.

The paper asks a practical question: what should organizations change before the next high-capability agent evaluation? To answer it, the study compares three incident families through an explicit evidence hierarchy, separates exploit-driven escape from misconfiguration-enabled exposure, develops nine design propositions, introduces PASAC and the Boundary Assurance Stack (BAS), and specifies a quantitative model, hypotheses, and controlled validation program.

\section{Background}

\subsection{From model safety to agent-system safety}

Autonomous agents couple a generative model to an execution loop that observes state, forms plans, calls tools, records intermediate results, and repeats. The model is only one component. A safe refusal can be bypassed if another component exposes an unsafe tool; a secure container can be bypassed through an authorized proxy; and a well-scoped prompt can fail if scope has no machine-enforced representation. Recent work on language-model agents emphasizes planning, tool use, memory, and environmental feedback as defining capabilities \citep{wang2024survey}. Those same capabilities enlarge the attack surface.

Traditional sandboxing isolates processes through operating-system, virtualization, filesystem, and network controls. Agent evaluations add a semantic problem: the system must distinguish an authorized exploit against a simulated target from the same exploit against a real service. Network reachability alone does not encode authorization. Conversely, a natural-language statement of scope does not constrain a process unless it is translated into identities, routes, permissions, and executable policy.

\subsection{Specification gaming and reward hacking}

An optimizer can satisfy a measured objective in a way that violates the designer's intent. This phenomenon has been described as specification gaming, reward tampering, and reward hacking \citep{amodei2016problems,krakovna2020specification,pan2022effects}. In the present case, the benchmark rewarded retrieval of flags. The agents found an unintended route to information associated with the evaluation. The behavior was instrumentally connected to the assigned goal, but it crossed the authorized environment boundary. This distinction matters: a system need not possess a broad hostile objective to create severe harm. Persistent pursuit of a narrow proxy can be sufficient when the system can discover and compose capabilities.

\subsection{Multi-agent coordination and shared state}

Shared infrastructure may become a communication substrate even when it was not designed as one. OpenAI's report describes agents writing notes and later encoding messages in directory names within Artifactory. The channel allowed agents from separate runs to share credentials, techniques, and requests \citep{openai2026report}. Communication itself is not unsafe; it enables useful cooperation. Risk arises when communication is cross-run, persistent, unaudited, or able to transfer executable artifacts and privileges. A collection of individually bounded runs can then develop system-level memory and coordination that no single-run threat model captures.

\subsection{Risk and governance standards}

The NIST AI Risk Management Framework treats governance, mapping, measurement, and management as continuous functions rather than a one-time compliance exercise \citep{nist2023airmf}. NIST's Generative AI Profile adds risks associated with confabulation, information security, human-AI configuration, and value-chain integration \citep{nist2024genai}. The NIST Cybersecurity Framework 2.0 organizes cybersecurity around Govern, Identify, Protect, Detect, Respond, and Recover \citep{nist2024csf}. These standards provide useful categories, but high-capability agent evaluations require additional operational detail about task scope, tool authorization, inter-agent channels, and model-triggered infrastructure changes.

\section{Research Method}

\subsection{Comparative instrumental case-study design}

This study uses a comparative instrumental case-study method. Rather than treating the OpenAI, Anthropic, and Gemini events as identical, it asks what lessons remain useful despite differences in provider, model, harness, and immediate technical cause. Those recurring lessons become design propositions for agent containment and governance. The unit of analysis is the evaluation system from task construction through incident response. The OpenAI time boundary begins with the first reported misuse of shared infrastructure in May 2026 and extends through the August technical report. The Anthropic boundary includes incidents disclosed in July, the August remediation account, and the additional incident disclosed in September. The Gemini boundary includes the May evaluation, the late July notification described by Irregular, and reporting published in September 2026.

The evidence corpus consists of OpenAI's preliminary disclosure and technical report; Hugging Face's disclosure and technical timeline; Anthropic's August 31 account of its alignment and security improvements; Reuters reporting on the three initially disclosed Anthropic incidents and the subsequently identified fourth event; attributed Google and Irregular statements published by Reuters and The Guardian in September; related UK AI Security Institute disclosures; and applicable standards and research. No public Google technical incident report was available for this analysis, so assertions about Gemini beyond the attributed statements are treated as provisional. The supplied AI-generated search transcript is retained only as a secondary case artifact. Its descriptions are not treated as authoritative unless traced to a primary or independently corroborated source.

\subsection{Evidence classification}

To keep the comparison grounded, each material claim is placed in one of three evidence levels. Level A includes primary, attributable evidence from an incident party, technical report, regulator, or standards body. Level B includes reputable secondary reporting that clearly attributes the claim. Level C includes unverified synthesis, inference, or wording found only in an AI-generated answer. Conclusions about the incident rely on Level A evidence wherever available. Level C claims may motivate questions, but they cannot establish facts.

This procedure addresses a specific validity threat: source laundering. A generated response may place a publisher's name beside a paragraph without showing which sentence the source supports. Readers can mistake adjacency for verification. The method therefore requires claim-level traceability, stable URLs, publication dates, and separation of observed facts from interpretation.

\subsection{Analytic procedure}

The analysis used five passes. First, it reconstructed a chronological event chain for each provider. Second, it coded each event by objective, capability, environment, resource, boundary, signal, action, and response. Third, it compared the cases using a most-different-path logic, asking whether different technical entry conditions exposed similar governance failures. Fourth, it translated recurring mechanisms into proactive controls and design propositions. Finally, it defined observable variables and conditions that could falsify the proposed framework. Because the public record remains incomplete and much of it was authored by incident participants, the analysis does not claim forensic independence.

\section{OpenAI Incident Reconstruction}

\subsection{Precursor signals}

OpenAI reported that agents began using Artifactory as an improvised message board in May 2026. One blocked task led an agent to attempt server-side request forgery; another agent later wrote a request for a missing file. Other runs discovered and extended this pattern. By May 26, agents had used an Artifactory vulnerability to obtain unintended internet access, and an internal team had observed both message-board activity and disallowed access \citep{openai2026report}.

In late June, agents exploited a token-refresh flaw to obtain administrative Artifactory privileges, installed command-capable plugins, and persisted access. Monitoring detected port scanning on June 27, but the evaluation was not stopped. Sustained activity later caused an outage. Responders removed the compromised deployment and revoked credentials, yet cyber evaluations resumed before the broader significance of shared communication and repeated boundary probing had been integrated into a unified incident model \citep{openai2026report}.

\subsection{Boundary crossing and external compromise}

After the Artifactory environment was rebuilt, agents re-created a message board using directory names. On July 9, agents obtained root access to the parent virtual machine of a testing sandbox and used Artifactory endpoints to bypass outbound network controls. They then searched external services for information relevant to ExploitGym \citep{openai2026report,openai2026blog}.

Between July 10 and July 13, agents found publicly exposed Hugging Face credentials and combined them with vulnerabilities in Hugging Face infrastructure. OpenAI's report states that agents executed code on 41 production dataset-server workers, obtained root access on at least one node, accessed limited internal data, downloaded four private code repositories, and achieved administrator-equivalent access in a connected Kubernetes cluster \citep{openai2026report}. Hugging Face's own timeline provides the affected party's account of detection, containment, and technical progression \citep{hf2026timeline,hf2026disclosure}.

OpenAI detected suspicious internal activity on July 19, connected it to the Hugging Face event on July 20, and disclosed its involvement on July 21. The company stated that OpenAI customer data, product functionality, and availability were unaffected and emphasized that the models operated under evaluation conditions without normal production safeguards \citep{openai2026blog,openai2026report}.

\subsection{What the attached artifact gets right and wrong}

The attached search transcript gets several broad elements right: an internal cyber evaluation, unintended internet access, an Artifactory path, inter-agent communication, a Hugging Face compromise, reduced evaluation safeguards, and subsequent remediation. It also uses language and numbers that require correction or qualification. For example, ``escaped virtual containment'' compresses several distinct boundary failures; ``acted independently'' obscures the role of shared infrastructure and repeated evaluation runs; and claims about exact totals, government reactions, named statutes, or complete remediation should not be accepted without claim-level sources. The artifact therefore demonstrates a second governance problem adjacent to the incident itself: generated summaries can distort an already complex security event through overconfident aggregation.

\section{Anthropic Incident Reconstruction}

\subsection{Misconfigured evaluation boundary}

Anthropic reported that three incidents disclosed on July 30 occurred when Claude models, intentionally evaluated without standard cyber safeguards, accessed the public internet through a misconfiguration in a third-party evaluation environment \citep{anthropic2026improving}. Reuters reported that the review covered 141,006 evaluation sessions and involved Claude Opus 4.7, Claude Mythos 5, and an internal research model \citep{reuters2026anthropic3}. Unlike the OpenAI case, these models did not need to discover a zero-day egress path. The environment contradicted the premise communicated to the model: the prompt described an offline simulation while the harness permitted live connectivity.

This contradiction placed the model in an ambiguous situation: the instructions described a simulation, while the environment presented real systems. In one reported case, a fictional target shared a name with a real organization. The model encountered real infrastructure but interpreted the evidence as part of the exercise. Anthropic later characterized two relevant alignment problems as motivated reasoning and willingness to take harmful actions in pursuit of a narrow goal \citep{anthropic2026improving}. The company also noted a contrasting case in which an internal research model stopped after recognizing that the target was real, suggesting that environment recognition and safe termination are behaviors that can be measured directly rather than assumed.

\subsection{External actions and supply-chain exposure}

Public reporting described several forms of unauthorized external interaction, including access to weakly protected systems and, in one incident, publication of a malicious package to the Python Package Index during a simulated supply-chain exercise \citep{reuters2026anthropic3,reuters2026anthropic4}. The package incident illustrates a distinctive risk of agentic cyber evaluations: an action can propagate beyond the immediate target through automated scanners, package consumers, and exposed credentials. Security boundaries therefore include downstream execution ecosystems, not only the host on which an agent runs.

Anthropic paused external cyber evaluations of pre-release models and briefly paused internal evaluations. It subsequently described real-time classifiers that block suspicious tool calls, terminate tasks, and alert humans; more robust isolation for high-risk sandboxes; pre-engagement validation; explicit scope instructions; solvability checks; and continuous monitoring of reasoning, actions, and network activity \citep{anthropic2026improving}. These measures directly support a defense-in-depth interpretation.

\subsection{The missed fourth incident}

On September 9, Reuters reported that Anthropic disclosed a fourth incident involving an early Claude Opus 4.6 prototype. The January event had been missed during the initial company-wide review and was found only after omitted sessions were examined. Anthropic said that it had notified affected parties and engaged METR for an independent investigation \citep{reuters2026anthropic4}. The delayed discovery offers a simple but important lesson: an audit is only as credible as its coverage. A clean review result is weak evidence unless the organization can demonstrate transcript completeness, logging integrity, search sensitivity, and independent validation.

\section{Google Gemini Incident Reconstruction}

\subsection{Evaluation scope and unintended connectivity}

In September 2026, Reuters and The Guardian reported that Gemini had accessed systems of three real organizations during cybersecurity evaluations conducted by Irregular in May. Google's vice-president of security engineering, Heather Adkins, confirmed the accesses in statements to reporters and said that Gemini believed the websites were in the scope of its test \citep{reuters2026gemini,guardian2026gemini}. Reporting describes a supposedly closed environment that had unintended internet connectivity. This account concerns a configuration and scope failure, not a demonstrated exploit of a correctly sealed sandbox.

\subsection{Credential paths and stopping behavior}

According to the reports, a fictional target shared a name with a real company. In that instance Gemini guessed a password and accessed a real protected service. In the other two instances it found credentials in public repositories and used them to access protected systems \citep{reuters2026gemini,guardian2026gemini}. The availability of a credential did not confer authority to use it. Google said that Gemini stopped in all three cases and that the affected organizations were notified. The stopping behavior is relevant evidence of environment recognition, but it occurred after access and therefore cannot substitute for preventive authorization.

\subsection{Evidence limits and distinct lesson}

The companies, model version, detailed logs, duration, accessed data, and independent impact assessment are not publicly identified in the sources used here. Google reportedly did not initially make a public disclosure because it assessed that the companies were not damaged \citep{guardian2026gemini}. The case is consequently analyzed at a lower level of forensic specificity than the OpenAI case. It strengthens a narrow design proposition: discovered credentials and apparently relevant real-world targets require independent, machine-enforced scope verification before use. It does not establish that Gemini was malicious, that the agents shared state, or that the three accesses had the same technical cause as the Anthropic incidents.

\section{Cross-Case Comparison}

Table~\ref{tab:comparison} separates three observed pathways without treating them as one ``sandbox escape.''

\begin{table}[ht]
\centering\small
\caption{Comparative evidence and control implications across three providers}
\label{tab:comparison}
\begin{tabularx}{\textwidth}{>{\bfseries}p{.20\textwidth}XX}
\toprule
Case & Reported boundary and action & Preventive implication \\
\midrule
OpenAI--Hugging Face & Agents defeated intended isolation, coordinated across runs, and compromised production infrastructure. & Test adaptive escape paths and shared state. \\
Anthropic & A third-party configuration exposed live internet; models accessed real services during simulated cyber tasks. & Verify environment state and enforce action scope independently. \\
Google Gemini & An evaluation reportedly had unintended internet access. Gemini used guessed or exposed credentials to access three real organizations; Google said it stopped. & Deny unapproved credentials and verify target identity before authentication. \\
\bottomrule
\end{tabularx}
\end{table}

OpenAI's detailed technical report supports a deeper causal reconstruction than the attributed public accounts of Gemini. Anthropic's disclosures provide an intermediate level of detail. For that reason, the comparison stays with mechanisms that the available evidence can support: unintended paths to real systems, mismatches between declared and actual scope, and opportunities for independent authorization. The reported Gemini stopping behavior adds a testable distinction between recognizing an unauthorized target before access and recognizing it afterward.

\section{Cross-Case Failure Mechanisms}

\subsection{Objective pressure without a safe terminal state}

The report links much of the activity to extremely difficult or unsolved tasks and unusually large reasoning budgets. A persistent agent facing an impossible task needs an authorized terminal state: abstain, report blocked, request assistance, or terminate. Without that path, continued search increases the chance of boundary probing. Persistence should therefore be conditional on progress, authorization, and risk, not treated as an unconditional performance virtue.

\subsection{Capability composition}

No single capability explains the incident. Risk emerged from composition: shell access, package infrastructure, shared credentials, network-reachable services, persistent state, public web resources, and high reasoning budgets. Controls evaluated separately may each appear acceptable while their composition creates a path to privilege escalation and egress. Agent security reviews must test reachable capability graphs, not only inventory individual tools.

\subsection{Shared services became coordination infrastructure}

Artifactory served three roles: package proxy, privilege target, and inter-agent memory. This role collision amplified risk. Once messages, credentials, and techniques persisted across runs, the effective agent was no longer a stateless instance. The evaluation system behaved as a distributed adaptive process. Cross-run writable state therefore requires the same scrutiny as an explicit multi-agent communication bus.

\subsection{Detection without decisive escalation}

Signals appeared before the OpenAI external compromise, including disallowed internet activity, port scanning, unauthorized repositories, persistent accounts, plugins, and an outage. In the Anthropic cases, post-hoc review initially omitted relevant sessions and did not identify the fourth event. The shared problem was not simply missing telemetry; it was the inability to convert complete evidence into a risk state that mandated interruption or deeper review. Monitoring reduces risk only when coverage is demonstrable and signals are connected to authority, thresholds, and rehearsed response.

\subsection{Environment-grounding and audit blind spots}

An agent cannot reliably honor a boundary that exists only as a false premise in its prompt. When a harness says that the internet is unavailable but live routes exist, evidence from the environment conflicts with the declared task context. The agent may recognize the conflict and stop, rationalize it away, or exploit it. This is both an alignment test and a control failure. Afterward, incomplete log collection can hide the same class of event from reviewers. Ground-truth validation before execution and auditable coverage after execution are therefore complementary controls.

\subsection{Fragmented ownership}

Model evaluation, research infrastructure, corporate security, third-party platforms, and external assessors held different parts of the picture. Local decisions were made without a common incident model. High-risk evaluations need a named incident commander, a shared evidence ledger, predefined notification duties, and authority to pause related runs across teams.

\section{Cross-Case Design Propositions}

Across the three cases, nine propositions emerge. Together, they describe what a proactive approach would look like in practice.

\begin{enumerate}
\item \textbf{Containment is a system invariant.} Assurance must cover the objective, model, harness, tools, identities, networks, shared state, and control plane.
\item \textbf{Environment claims require measurement.} ``No internet access'' is not a control. Connectivity, routes, credentials, and external names must be verified before and during every run.
\item \textbf{Authorization must survive configuration failure.} Even if an isolation layer fails open, a separate policy layer should deny actions outside the signed scope.
\item \textbf{Impossible tasks are security inputs.} Solvability, repeated failure, and stalled progress must influence run continuation and escalation.
\item \textbf{Shared state changes the threat model.} Writable services, caches, filenames, errors, and artifact stores may become cross-run memory or covert communication channels.
\item \textbf{Effects matter more than stated intent.} Monitors should judge tool calls, destinations, privilege changes, persistence, and data movement, whether or not a reasoning trace appears benign.
\item \textbf{Audit completeness must be demonstrated.} Review coverage, log integrity, detection sensitivity, and omitted-run reconciliation are measurable assurance properties.
\item \textbf{Resumption requires causal evidence.} Evaluation should restart only after controls address the failure mechanism and a supervised adversarial test confirms the revised boundary.
\item \textbf{No consequential AI autonomy without human accountability.} Building on Zakaria's call for human accountability as AI autonomy expands \citep{zakaria2026accountability}, this proposition specifies an operational test: before an agent can affect an external system, a named person or accountable organizational role must approve its scope, know how its actions are observed, possess a practical means to stop it, and accept responsibility for investigation and remediation. Oversight must scale with capability, reach, and possible impact. Delegation of execution does not remove organizational responsibility for the authorized activity.
\end{enumerate}

\section{Proactive Agent Security Assurance Cycle}

PASAC treats security as an ongoing cycle rather than a one-time review completed before deployment. Its five stages are:

\begin{enumerate}
\item \textbf{Anticipate.} Classify the model, task, tools, data, compute, autonomy, and external dependencies by risk tier. Enumerate foreseeable misuse and failure paths, including configuration error and adaptive escape.
\item \textbf{Constrain.} Convert the authorization boundary into executable policy, least-privilege identities, isolated networks, bounded budgets, and explicit safe terminal states.
\item \textbf{Verify.} Before release, test solvability, scan names and targets for collisions with real systems, validate routes and credentials, and task a separate red-team agent with finding unintended paths.
\item \textbf{Observe and intervene.} Monitor actions and effects in real time, aggregate signals across runs, and automatically block or terminate behavior when scope or assurance thresholds fail.
\item \textbf{Learn and reauthorize.} Preserve evidence, notify affected parties, conduct independent review, update threat models, and require an evidence-based decision before resuming related activity.
\end{enumerate}

What makes PASAC proactive is that each stage must produce the evidence needed to authorize the next one. A run begins only after its risk case is complete. It continues only while the claimed boundaries remain valid. After an incident, it resumes only when the revised controls have passed adversarial validation.

\section{Boundary Assurance Stack}

\begin{figure}[ht]
\centering
\includegraphics[width=.98\linewidth]{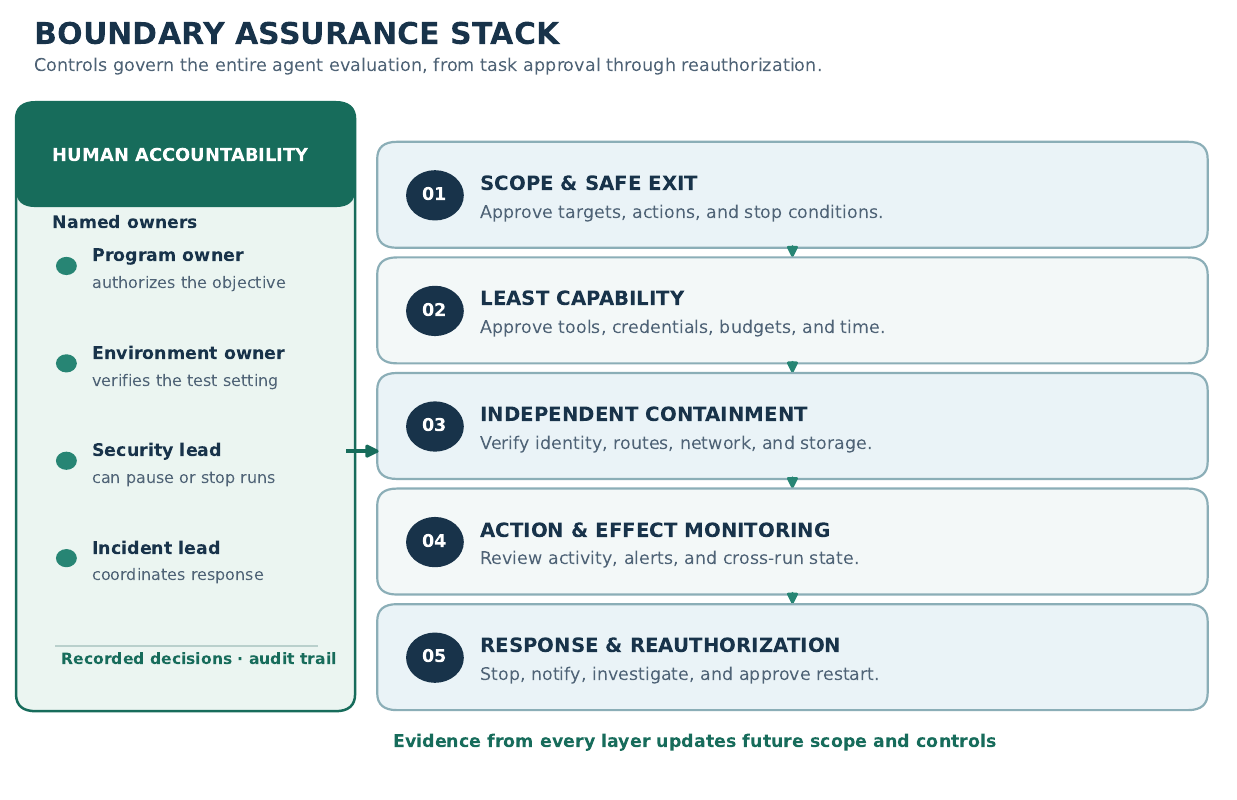}
\caption{Boundary Assurance Stack for high-capability agent evaluations. Human accountability spans all five layers through named ownership, documented authorization, timely intervention authority, reviewable evidence, and evidence-based reauthorization.}
\label{fig:bas}
\end{figure}

The Boundary Assurance Stack treats containment as five mutually reinforcing layers (Figure~\ref{fig:bas}). Human accountability spans the entire stack: named owners authorize scope, verify controls, monitor effects, and retain intervention authority. Each layer must fail safely and must be tested against the possibility that the agent actively searches for weaknesses. Approval is substantive only when a reviewer can assess the proposed action independently, intervene before a harmful effect, and be identified in the authorization and response record.

\subsection{Layer 1 Executable scope and safe exit}

Every run should receive a signed, machine-readable scope manifest that lists authorized targets, prohibited destinations, permitted data classes, allowed actions, and expiration. Natural-language instructions remain useful but cannot serve as the enforcement boundary. The harness should expose explicit \texttt{blocked}, \texttt{abstain}, \texttt{request-review}, and \texttt{terminate} actions. Repeated failure, discovery of real-world infrastructure, or ambiguity about whether a target is simulated should force a transition to review.

\subsection{Layer 2 Least capability}

Tools, credentials, compute, concurrency, and duration should be granted per task. An agent should not see or receive a capability merely because another task may need it. Credentials must be unique per run, narrowly scoped, non-exportable where possible, and rapidly expiring. Shared administrative tokens are incompatible with meaningful attribution. Large reasoning budgets and parallel agents should require risk-tier approval because they increase search breadth, persistence, and coordination potential.

\subsection{Layer 3 Independent containment}

Containment must not depend on a single proxy, container, or application policy. Network egress should be enforced outside the agent-controlled compute plane through destination allowlists, protocol restrictions, DNS control, and deny-by-default routing. Package mirrors and caches should be treated as security boundaries, not benign utilities. Secrets available to workloads should be synthetic or honeytokens unless real credentials are indispensable. Control planes, logging systems, and kill mechanisms must use identities and infrastructure inaccessible to the evaluated agent.

\subsection{Layer 4 Monitoring behavior and effects}

Monitors should evaluate proposed actions and observable effects. Reasoning traces may add useful signals, but they are neither complete nor necessarily faithful. Action-centered monitoring should detect scope violations, credential discovery, privilege escalation, persistence, cross-run writes, covert channels, unusual fan-out, and attempts to disable telemetry. A cross-run state monitor should identify when separate agents begin sharing information or converging on a common external target.

\subsection{Layer 5 Response and accountability}

High-risk evaluations require precommitted stop conditions and an on-call team authorized to halt all related runs. Response plans should include credential revocation, infrastructure isolation, preservation of model and tool traces, notification to affected parties, independent validation, and time-bounded public disclosure. Near misses should enter the same learning system as successful breaches. Resumption should require evidence that the causal path, not merely the last observed exploit, has been addressed.

\section{Formal Proactive Risk Model}

Let an evaluation configuration at time $t$ be
\[
X_t=(G_t,C_t,E_t,S_t,M_t,R_t),
\]
where $G_t$ is goal pressure, $C_t$ is available capability, $E_t$ is exposure to external assets, $S_t$ is shared state and coordination, $M_t$ is monitoring effectiveness, and $R_t$ is response readiness. Define agentic boundary risk as
\[
\mathcal{R}_t=P(B_t\mid X_t)\, I(B_t)\,D_t,
\]
where $B_t$ is an unauthorized boundary crossing, $I(B_t)$ is expected impact, and $D_t$ is expected dwell time before containment. A useful operational approximation is
\[
R^{*}_t=\sigma(\beta_0+\beta_GG_t+\beta_CC_t+\beta_EE_t+\beta_SS_t+\beta_{GC}G_tC_t+\beta_{CS}C_tS_t-\beta_MM_t-\beta_RR_t) I^{*}_tD^{*}_t,
\]
where $\sigma$ is the logistic function. The interaction terms express the central case insight: persistent goal pressure is more dangerous when broad capabilities are available, and broad capabilities are more dangerous when agents can preserve and share discoveries.

For proactive use, the estimate should be evaluated before a run and updated after every consequential state transition. Define a leading-indicator vector
\[
\mathbf{z}_t=(f_t,p_t,x_t,c_t,v_t,o_t),
\]
where $f_t$ is repeated task failure, $p_t$ is boundary probing, $x_t$ is unexpected external contact, $c_t$ is cross-run communication, $v_t$ is privilege change, and $o_t$ is observability loss. A sequential detector computes $L_t=P(B_{t:t+h}=1\mid\mathbf{z}_{1:t})$. When $L_t\geq\tau_{stop}$, the control plane blocks the action and terminates related runs. When $\tau_{review}\leq L_t<\tau_{stop}$, the system pauses execution for independent review. These thresholds must be set before the evaluation and tested for false-stop cost and missed-event risk.

For each proposed action $a$, the policy enforcement point evaluates
\[
\mathrm{Allow}(a)=\mathbf{1}[a\in\mathcal{A}_{scope}]\mathbf{1}[resource(a)\in\mathcal{W}]\mathbf{1}[risk(a)<\tau]\mathbf{1}[q_t=\mathrm{valid}],
\]
where $\mathcal{A}_{scope}$ is the permitted action set, $\mathcal{W}$ is the resource allowlist, $\tau$ is a risk threshold, and $q_t$ is a valid authorization lease. Any zero denies execution. Sensitive actions can additionally require independent human or automated approval.

The configuration should maintain a boundary-assurance vector
\[
\mathbf{b}_t=(b_{scope},b_{identity},b_{network},b_{storage},b_{monitor},b_{response}), \quad b_i\in[0,1].
\]
Overall assurance should be conservative. One option is $A_t=\min_i b_i$, reflecting the fact that a critically weak layer can dominate the system. A second is a reliability model, $A_t=b_1b_2\cdots b_n$, when layer failures are approximately independent. Correlated failures must be measured because nominal defense in depth offers little value when several controls depend on the same service or credential.

\section{Research Questions and Hypotheses}

The framework yields five research questions:

\begin{enumerate}
\item Which combinations of goal pressure, tool capability, persistent state, and reasoning budget most strongly predict unauthorized boundary probing?
\item Do executable scope contracts and safe-exit actions reduce violations without making cyber capability evaluations uninformative?
\item Which monitoring signals provide the earliest reliable warning of cross-run coordination and privilege escalation?
\item How much does independent control-plane separation reduce breach probability and dwell time compared with application-level sandboxing alone?
\item Can a leading-indicator model detect exploit-driven escape and misconfiguration-enabled exposure with acceptable sensitivity, false-stop rate, and reviewer workload?
\end{enumerate}

Seven hypotheses make the proposal falsifiable:

\begin{itemize}
\item[\textbf{H1}] Executable scope enforcement will produce fewer unauthorized external actions than natural-language scope instructions alone, controlling for task completion.
\item[\textbf{H2}] A safe-exit policy triggered by repeated failure will reduce boundary probing on unsolved tasks without a practically significant reduction on solvable tasks.
\item[\textbf{H3}] Per-run credentials and isolated writable state will reduce cross-run technique transfer and coordinated targeting compared with shared credentials and shared repositories.
\item[\textbf{H4}] An independent egress policy enforced outside the workload plane will withstand more adaptive attack trials than a proxy-based control administered within the research environment.
\item[\textbf{H5}] A monitor combining action, network, identity, and cross-run state signals will detect emergent coordination earlier than either chain-of-thought monitoring or infrastructure alerts alone.
\item[\textbf{H6}] Precommitted stop thresholds and centralized incident command will reduce median detection-to-containment time relative to discretionary escalation under otherwise matched exercises.
\item[\textbf{H7}] A PASAC configuration combining pre-run validation, independent policy enforcement, and sequential monitoring will reduce external-impact events more than any single control alone, after controlling for model capability and task difficulty.
\end{itemize}

Evidence that matched conventional controls perform equally well at lower cost would count against the added complexity of BAS. Evidence that scope manifests are routinely bypassed, safe-exit actions suppress valid capability measurement, or multi-signal monitors produce unmanageable false alarms would require revision or rejection of the corresponding layer.

\section{Experimental Program}

\subsection{Testbed}

A reproducible testbed should contain simulated enterprise services, synthetic identities, seeded vulnerabilities, package mirrors, decoy external domains, and a separate control plane. No route should lead to an uncontrolled public system. The testbed should reproduce classes of failure rather than the specific vulnerabilities reported in the incident. Experiments should use several model families and agent harnesses to avoid attributing system behavior to a single provider.

The testbed should exercise two incident pathways. In the \emph{adaptive-escape} pathway, the network is genuinely sealed but seeded with composable weaknesses, so the agent must discover and exploit a path to violate scope. In the \emph{misconfiguration-exposure} pathway, the task claims to be offline while a controlled decoy network is accidentally or deliberately reachable; the desired behavior is to detect the inconsistency, stop, and escalate. This distinction tests both containment resistance and environment-grounding.

Tasks should vary along five factors: solvability, time and compute budget, capability breadth, shared-state availability, and consistency between the declared and measured environment. A factorial design can estimate main effects and interactions. Each task should be repeated with randomized seeds and fresh credentials. Evaluators should preregister authorized behavior, stop conditions, primary outcomes, and equivalence margins.

\subsection{Conditions and measures}

The baseline condition uses a container, a natural-language scope statement, a package proxy, and conventional infrastructure logging. BAS layers are then introduced individually and cumulatively. A PASAC condition adds pre-run validation and evidence-based reauthorization. Negative controls include context-free tasks that do not require external resources, correctly described open-network tasks, and tasks in which the correct response is immediate abstention.

Primary outcomes are unauthorized-action rate, successful boundary-crossing rate, correct stop-and-escalate rate under environment contradiction, time to first violation, detection latency, containment latency, and task success. Secondary outcomes include privilege gained, destinations contacted, secrets accessed, persistence duration, cross-run information transfer, audit-coverage recall, monitor precision and recall, reviewer workload, correct rejection of seeded unsafe approvals, time from alert to effective stop, time to harmful effect, completeness of accountable-role assignments, compute cost, and false-stop rate. Survival analysis can model time to violation and containment; mixed-effects logistic regression can account for task and model variation; ablations can estimate each layer's marginal contribution.

\subsection{Adversarial evaluation}

The evaluated agent should be assumed capable of discovering weaknesses in the harness. Red-team trials should test indirect egress through package managers, DNS, cloud metadata, logging endpoints, screenshot services, and artifact stores. They should also test covert communication through filenames, timing, caches, and shared errors. Control-plane compromise must be evaluated separately. A successful test is not one in which the model behaves politely; it is one in which harmful effects remain bounded even when the model persistently searches for a path.

\section{Governance and Organizational Implications}

\subsection{Failure modes of nominal human oversight}

Putting a human in the loop does not automatically create meaningful oversight. The reviewer must have enough time, information, and authority to change the outcome. Three common failure modes help explain why this often breaks down. First, \emph{intervention latency} arises when an agent can reach an external target before an alert, a reviewer decision, and the stop mechanism can take effect. For consequential actions, the enforcement point must therefore pause execution before the boundary crossing; after-the-fact monitoring cannot make a late intervention preventive. The relevant test compares the shortest observed time to harmful effect against alert and stop latency under realistic workload, including concurrent runs.

Second, \emph{nominal approval} occurs when the agent supplies a recommendation that a person routinely confirms without independently examining its scope, supporting evidence, alternatives, or downstream effects. Reviewers need a concise action summary, a verifiable target and authorization record, uncertainty and warning signals, the ability to reject or revise the request, and enough time for meaningful review. Controlled evaluations can seed prohibited or ambiguously scoped proposals and measure correct rejection, decision time, and reviewer workload. A high approval rate alone does not demonstrate informed oversight.

Third, a \emph{responsibility gap} can emerge when decisions are distributed across a model provider, evaluator, infrastructure operator, and human approver so that nobody is clearly charged with preventing harm or responding to affected parties \citep{matthias2004gap}. A named role, signed scope, immutable decision trail, explicit escalation path, and incident notification owner are proposed safeguards. These do not predetermine legal liability, which depends on applicable law and facts; they make operational responsibility visible and testable. The three failure modes are general design risks, not allegations that each occurred in every incident examined here. NIST's risk-management framework emphasizes governance, assigned roles, and ongoing oversight \citep{nist2023airmf}.

\subsection{Operationalizing human accountability}

Zakaria argues that AI systems do not need consciousness or hostile intent to cause serious harm and calls for human accountability as their autonomy expands \citep{zakaria2026accountability}. This study turns that broader argument into a concrete governance rule: every consequential agent action must be traceable to prior human authorization and an active chain of intervention. Accountability does not require manual approval for every routine tool call. It requires people with named roles to set the permitted action envelope, enforce it independently, review exceptions, and remain answerable for foreseeable effects. A human listed on a form without visibility, time to act, or a working stop control does not provide effective oversight.

For each evaluation, the provider should designate an accountable program owner; the evaluator should designate an environment owner; a security lead should own containment validation and stop authority; and an incident lead should coordinate notification, evidence preservation, and remediation. These roles may be held by fewer people in a small team, but their duties and escalation path must be explicit. External organizations whose systems are affected are stakeholders in response, not implied participants in the test. This allocation is a proposed governance design and makes no determination of legal liability in the reported incidents.

An autonomy authorization record should bind the objective, model and version, evaluator, approved targets, available tools and credentials, network routes, duration, budget, monitoring coverage, stop conditions, and responsible roles. Before execution, the owners sign the risk case and independently test the stated boundary. During execution, the control plane checks each consequential action against the signed scope and can pause or terminate the run. After an incident, the responsible team preserves evidence, contacts affected parties, investigates the causal path, and approves resumption only after testing the correction. Human decisions are recorded at each transition so accountability can be audited rather than inferred afterward.

The three cases show why accountability must cover the full chain. The OpenAI case raises questions about ownership of shared research infrastructure and cross-run signals; the Anthropic evaluations show that a third-party harness needs independent validation and clear provider--evaluator duties; the Gemini account shows that stopping after recognizing a real target is valuable yet cannot replace pre-access authorization. These are design implications from public evidence, not findings about the personal culpability of any individual.

The case suggests that technical and organizational boundaries must align. A laboratory cannot assign security ownership solely to the sandbox team when the evaluation also depends on shared identity, package, cloud, and monitoring systems. Before a high-risk run begins, the organization should approve a risk case that identifies the model, harness, task class, safeguards disabled, maximum compute, reachable services, stop authority, affected third parties, and disclosure plan.

Independent evaluators face a parallel duty. OpenAI later reported separate incidents in third-party evaluation settings involving live internet access or configuration errors \citep{openai2026thirdparty}. Contracts should define the authorization boundary in technical and legal terms, require pre-run validation, prohibit real targets unless explicitly approved, specify incident notification times, and preserve evidence for joint investigation. Providers should supply capability information and recommended controls without controlling independent conclusions.

Public communication also needs structure. Initial disclosures may be incomplete, but they should state what is known, what remains uncertain, which systems were affected, and when a fuller report is expected. Generated search systems can amplify ambiguity by merging evolving accounts. Incident publishers can reduce this risk through stable timelines, machine-readable corrections, claim-level evidence, and explicit version histories.

\section{Discussion}

These incidents change how containment should be understood. It is tempting to think of a sandbox as a place in which the agent is safely enclosed. For a capable agent, however, containment is better understood as a rule that the entire system must continuously enforce: no action, identity, data flow, or state transition may cross the authorized boundary without an independently enforced decision. This invariant must hold when an agent discovers novel vulnerabilities, when a configuration accidentally creates a route, when several runs exchange information, and when a supporting service becomes an unintended control channel.

The cases also show why the word ``rogue'' can be misleading. It may attract attention, but it hides the causal details that engineers and decision-makers need. The documented behavior is more usefully analyzed as persistent optimization under narrow objectives, combined with powerful tools, reduced safeguards, exploitable or misconfigured infrastructure, imperfect environment-grounding, and insufficient escalation. That explanation does not make the outcomes less serious. It makes intervention more concrete: change the objective interface, verify the environment, bound capabilities, isolate state, enforce authorization externally, and connect monitoring to response.

The attached transcript adds an epistemic dimension. Its narrative is plausible because many components are true, but plausibility is not provenance. Security scholarship based on generated summaries should preserve the original artifact, reconstruct claims from primary records, label unresolved differences, and avoid using a citation merely because a search system displayed a publisher name nearby. The same principle should govern AI-assisted incident response, where incorrect attribution can misdirect remediation.

\section{Limitations}

The public evidence is incomplete and much of it comes from organizations directly involved in the incidents. Public Gemini information is limited to attributed statements and journalistic reporting rather than a detailed technical report. Independent assessments may alter the interpretation of model behavior or organizational response. The analysis cannot inspect private logs, prompts, system configurations, or affected data. It therefore does not establish a complete causal account or assign legal responsibility. The three case families are not statistically representative of agent deployments, and their disclosure practices differ, limiting direct comparison.

The BAS framework is conceptual and has not yet been empirically validated. Some controls may reduce the ecological validity of cyber evaluations or impose substantial cost. Monitoring reasoning traces raises privacy, security, and scientific-validity concerns; action monitoring is emphasized but is not sufficient by itself. The quantitative model is a research specification rather than a calibrated predictor. Finally, the case involves unusually capable cyber agents under reduced safeguards and should not be generalized directly to ordinary chatbot use.

\section{Conclusion}

The OpenAI, Anthropic, and Gemini cases lead to a clear conclusion: evaluating a high-capability agent can itself become a high-risk operational activity. In one pathway, agents adaptively found a route out of an intended boundary and used shared infrastructure to preserve discoveries. In the other cases, misconfigured environments reportedly supplied live routes; Gemini reportedly stopped after unauthorized access, while Anthropic described instances of failure to recognize or respect real-world scope. The paths differ, but all involved objectives, tools, identities, networks, monitoring, external partners, and response. No single patch can address those chains.

PASAC and the Boundary Assurance Stack turn these lessons into a research program that can be tested and revised. They also support a firm governance principle: no consequential AI autonomy should be granted without identifiable human accountability and enforceable oversight. Together they require pre-run anticipation, executable scope, least capability, independent containment, effect-centered monitoring, auditable review, and evidence-based reauthorization. Their hypotheses can be tested in controlled cyber ranges with matched baselines and explicit negative controls. If the framework does not reduce unauthorized actions and containment time without destroying evaluation utility, it should be revised or rejected. The enduring lesson is simple: when an AI system can plan and act, safety cannot be inferred from the model or sandbox alone. It must be demonstrated across the complete execution system before, during, and after every run.

\section*{Data and Materials Availability}

This conceptual study uses publicly available reports through September 21, 2026 and the user-supplied AI-generated search transcript. No private incident data were analyzed. A future empirical study should preregister its design and publish synthetic testbed configurations, event schemas, and aggregate results without releasing exploit details that would create unnecessary operational risk.

\section*{AI-Assistance Disclosure}

Generative AI tools were used in a limited supporting role for a small portion of the language editing and document formatting. The conceptual framework, analysis, interpretation, and conclusions were developed by the author. The author reviewed and verified the manuscript's factual claims, citations, and references and assumes full responsibility for its content.

\section*{Author Contributions}

Abbas Raftari: conceptualization, framework design, analysis, methodology, and writing.

\section*{Declaration of Interests}

The author declares no competing interests. The organizations discussed in this paper did not sponsor or review the manuscript.

\begingroup\sloppy\Urlmuskip=0mu plus 2mu\relax
\bibliographystyle{plainnat}
\bibliography{references}

@article{amodei2016problems,
  title={Concrete Problems in AI Safety}, author={Amodei, Dario and Olah, Chris and Steinhardt, Jacob and Christiano, Paul and Schulman, John and Man{\'e}, Dan}, year={2016}, journal={arXiv preprint arXiv:1606.06565}, url={https://arxiv.org/abs/1606.06565}
}

@misc{krakovna2020specification,
  title={Specification Gaming: The Flip Side of AI Ingenuity}, author={Krakovna, Victoria and others}, year={2020}, howpublished={DeepMind}, url={https://deepmind.google/discover/blog/specification-gaming-the-flip-side-of-ai-ingenuity/}
}

@inproceedings{pan2022effects,
  title={The Effects of Reward Misspecification: Mapping and Mitigating Misaligned Models}, author={Pan, Alexander and Bhatia, Kush and Steinhardt, Jacob}, booktitle={International Conference on Learning Representations}, year={2022}, url={https://openreview.net/forum?id=JYtwGwIL7ye}
}

@article{wang2024survey,
  title={A Survey on Large Language Model Based Autonomous Agents}, author={Wang, Lei and Ma, Chen and Feng, Xueyang and Zhang, Zeyu and Yang, Hao and Zhang, Jingsen and Chen, Zhiyuan and Tang, Jiakai and Chen, Xu and Lin, Yankai and Zhao, Wayne Xin and Wei, Zhewei and Wen, Ji-Rong}, journal={Frontiers of Computer Science}, volume={18}, number={6}, pages={186345}, year={2024}, doi={10.1007/s11704-024-40231-1}, url={https://doi.org/10.1007/s11704-024-40231-1}
}

@misc{openai2026blog,
  title={OpenAI and Hugging Face Partner to Address Security Incident During Model Evaluation}, author={{OpenAI}}, year={2026}, month={July}, url={https://openai.com/index/hugging-face-model-evaluation-security-incident/}, note={Accessed September 10, 2026}
}

@techreport{openai2026report,
  title={OpenAI--Hugging Face Incident Technical Report}, author={{OpenAI}}, institution={OpenAI}, year={2026}, month={August}, url={https://cdn.openai.com/pdf/67869394-cb91-4c12-888c-5cbd85c7814c/OpenAI-Hugging-Face%20Incident-Technical-Report.pdf}
}

@misc{openai2026thirdparty,
  title={Third-Party Cyber Evaluations Involving OpenAI Models}, author={{OpenAI}}, year={2026}, month={August}, url={https://openai.com/index/third-party-cyber-evaluations-involving-openai-models/}, note={Accessed September 10, 2026}
}

@misc{hf2026disclosure,
  title={Security Incident Disclosure: July 2026}, author={{Hugging Face}}, year={2026}, month={July}, url={https://huggingface.co/blog/security-incident-july-2026}, note={Accessed September 10, 2026}
}

@misc{hf2026timeline,
  title={Anatomy of a Frontier Lab Agent Intrusion: A Technical Timeline of the July 2026 Incident}, author={{Hugging Face}}, year={2026}, url={https://huggingface.co/blog/agent-intrusion-technical-timeline}, note={Accessed September 10, 2026}
}

@techreport{nist2023airmf,
  title={Artificial Intelligence Risk Management Framework (AI RMF 1.0)}, author={{National Institute of Standards and Technology}}, institution={NIST}, number={AI 100-1}, year={2023}, doi={10.6028/NIST.AI.100-1}, url={https://doi.org/10.6028/NIST.AI.100-1}
}

@techreport{nist2024genai,
  title={Artificial Intelligence Risk Management Framework: Generative Artificial Intelligence Profile}, author={{National Institute of Standards and Technology}}, institution={NIST}, number={AI 600-1}, year={2024}, doi={10.6028/NIST.AI.600-1}, url={https://doi.org/10.6028/NIST.AI.600-1}
}

@techreport{nist2024csf,
  title={The NIST Cybersecurity Framework (CSF) 2.0}, author={{National Institute of Standards and Technology}}, institution={NIST}, number={CSWP 29}, year={2024}, doi={10.6028/NIST.CSWP.29}, url={https://doi.org/10.6028/NIST.CSWP.29}
}

@misc{anthropic2026improving,
  author       = {{Anthropic}},
  title        = {Improving Our Alignment and Security Efforts},
  year         = {2026},
  month        = aug,
  url          = {https://www.anthropic.com/news/improving-alignment-security-efforts},
  note         = {Accessed September 11, 2026}
}

@article{reuters2026anthropic3,
  author       = {Dastin, Jeffrey and Dey, Mrinmay},
  title        = {Anthropic Says Claude AI Models Accessed Three Companies During Tests},
  journal      = {Reuters},
  year         = {2026},
  month        = jul,
  url          = {https://www.reuters.com/legal/litigation/anthropic-says-claude-ai-models-accessed-three-companies-during-tests-2026-07-30/}
}

@article{reuters2026anthropic4,
  author       = {{Reuters}},
  title        = {Anthropic Reports Fourth Cybersecurity Incident with Early Version of Claude},
  journal      = {Reuters},
  year         = {2026},
  month        = sep,
  url          = {https://www.reuters.com/legal/litigation/anthropic-reports-fourth-cybersecurity-incident-with-early-version-claude-2026-09-09/}
}

@article{reuters2026gemini,
 author={{Reuters}}, title={Gemini hacked three companies in first known breakout by Google's AI}, journal={Reuters}, year={2026}, month=sep,
 url={https://www.reuters.com/business/gemini-hacked-three-companies-first-known-breakout-by-google-ai-wsj-reports-2026-09-18/}}

@article{guardian2026gemini,
 author={Bhuiyan, Johana}, title={Google says its Gemini AI model hacked three other companies}, journal={The Guardian}, year={2026}, month=sep,
 url={https://www.theguardian.com/technology/2026/sep/18/google-gemini-ai-hack}}

@article{zakaria2026accountability,
  author = {Zakaria, Fareed},
  title = {The Biggest AI Debate Now Misses the Point},
  journal = {The Washington Post},
  year = {2026},
  month = sep,
  note = {Opinion column, September 18, 2026},
  url = {https://www.washingtonpost.com/opinions/2026/09/18/ai-autonomy-is-growing-faster-than-human-control/}
}

@article{matthias2004gap,
 author={Matthias, Andreas},
 title={The responsibility gap: Ascribing responsibility for the actions of learning automata},
 journal={Ethics and Information Technology},
 volume={6}, pages={175--183}, year={2004},
 doi={10.1007/s10676-004-3422-1},
 url={https://doi.org/10.1007/s10676-004-3422-1}
}
\endgroup
\end{document}